Title: Direct Numerical Simulation Tests of Eddy Viscosity in Two
Dimensions

\documentstyle[preprint,aps]{revtex}

\begin{document}
\draft
\title{Direct numerical simulation tests \\
of eddy viscosity in two dimensions}
\author{A. Chekhlov and S.A. Orszag}
\address{
Program in Applied and Computational Mathematics, Princeton University, \\
Princeton, NJ 08544}
\author{S. Sukoriansky}
\address{
Department of Mechanical Engineering, Ben-Gurion University of the Negev,\\
Beer-Sheva 84105, Israel}
\author{B. Galperin}
\address{
Department of Marine Science, University of South Florida,\\
St. Petersburg, FL 33701}
\author{I. Staroselsky}
\address{Cambridge Hydrodynamics, Inc., P.O. Box 1403, Princeton, NJ 08542}
\date{\today}
\maketitle
\begin{abstract}
Two-parametric eddy viscosity (TPEV) and other spectral
characteristics of two-dimensional (2D) turbulence in the energy
transfer sub-range are calculated from direct numerical simulation (DNS)
with 512$^2$ resolution. The DNS-based TPEV is
compared with those calculated from the test field model (TFM) and from the
renormalization group (RG) theory. Very good agreement between all
three results is observed.
\end{abstract}
\pacs {}

\narrowtext

\label{sec:level1}

Two-dimensional incompressible turbulent flows are described by the
vorticity equation:
\begin{equation}
{\partial \zeta \over \partial t} + {\partial \left (\nabla^{-2} \zeta,
\zeta \right ) \over \partial (x, y) }
= \nu_0 \nabla^2 \zeta, \label{one}
\end{equation}
where $\zeta$ is fluid vorticity and $\nu_0$ is molecular viscosity.
It is well known that the existence of inviscid invariants
$\int d^2x \zeta ^{2n}$ of (\ref{one}) results
in the flux of energy towards the largest spatial scales.
The presence of this inverse cascade complicates the large-scale
description of 2D flows and requires refinement of the
classical hydrodynamic notion of ``eddy viscosity.''
The concept of eddy viscosity is well defined for 3D turbulent flows,
where energy  cascades towards the smallest flow scales where it is
dissipated. To achieve an adequate coarse-grained description of 3D flow,
one can introduce increased ``effective'' dissipation at large scales
which accounts for the unresolved dissipation.

In 2D flows, the inverse flux of energy at large scales and
enstrophy dissipation at small scales make the eddy viscosity concept
more subtle. It was argued by  Kraichnan \cite{Kraichnan_76} that,
in Fourier space, a 2D eddy viscosity should include two parameters:
a cutoff wave number $k_c$ (which essentially
determines the size of the coarse grain), and the wave number of a given
mode, $k$.  The two-parameter eddy viscosity (TPEV), denoted
by $\nu(k\vert k_c)$, describes the energy exchange between a given resolved
vorticity mode with the wave number $k$ and all subgrid, or unresolved, modes
with $k > k_c$; it provides correct account for the energy and enstrophy
fluxes between resolved and unresolved scales.  The TPEV is derived from
the evolution equation for the spectral enstrophy density
$\Omega({k},t)\equiv\pi k\langle\zeta({\bf k},t)\zeta(-{\bf k},t)\rangle$,
where $\langle\ldots\rangle$ denotes averaging over thin circular shells:
\begin{eqnarray}
\left(\partial_t+ 2 \nu k^2\right) \Omega({k},t) = T_{\Omega}(k,t). \label{two}
\end{eqnarray}
Here, the enstrophy transfer function $T_{\Omega}(k,t)$ is given by
\begin{equation}
T_{\Omega}(k,t)=\pi k\int_{{\bf p}+{\bf q}={\bf k}}
\frac{{\bf p} \times {\bf q}}{p^2}
\langle\zeta({\bf p},t)\zeta({\bf q},t)\zeta(-{\bf k},t)\rangle
\frac{d{\bf p}~d{\bf q}}{(2\pi)^2} \ + \ {\rm c.c.}, \label{three}
\end{equation}
where c.c. stands for the complex conjugate term.
Assuming that the system is in statistical steady state and extending
integration in (\ref{three}) only over all such triangles $({\bf k, ~p, ~q})$
that $\vert k-p \vert < q < k+p$ and $p$ and/or $q$ are greater than $k_c$,
one defines the two-parametric transfer $T_{\Omega}(k\vert k_c)$ and
TPEV \cite{Kraichnan_76}:
\begin{eqnarray}
\nu (k\vert k_c)= - {T_{\Omega}(k\vert k_c)\over 2 k^2 \Omega(k)}. \label{four}
\end{eqnarray}

In a wide class of quasi-normal approximations \cite{mccomb} the
two-parametric transfer $T_{\Omega}(k\vert k_c)$ in two dimensions is given by
\begin{eqnarray}
T_{\Omega}(k \vert k_c)&&=\int \!\int_\Delta \Theta_{-k,p,q} (p^2-q^2)
\sin\alpha \left [{p^2-q^2 \over {p^2 q^2}} \Omega(p) \Omega(q) \right .
\nonumber \\
&&- \left . {k^2 - q^2 \over {k^2 q^2} } \Omega(q) \Omega(k) + {k^2 -
p^2 \over {k^2 p^2}} \Omega(p) \Omega(k) \right ]~dp dq,  \label{five}
\end{eqnarray}
where $\Theta_{-k,p,q}$ is the triad relaxation time.
Here, the angle $\alpha$ is formed by the vectors ${\bf p}$ and ${\bf q}$,
and $\int \!\int_\Delta$ denotes integration over the area defined above
(\ref{four}).

The main difference between various spectral closure models is in specification
of $\Theta_{-k,p,q}$.  In \cite{Kraichnan_76}, $T_{\Omega}(k \vert k_c)$ was
evaluated using TFM.  It was found that TPEV is a
sign-changing function of the form $\nu (k \vert k_c)= \vert \nu (0\vert k_c)
\vert N(k/k_c)$, with $\nu (0\vert k_c) < 0$, $N(0)=-1$, and $N(1)\approx 2.1$.
The derivation of $\Theta_{-k,p,q}$ using the RG theory was given in
\cite{dannevik} and adapted for 2D isotropic and anisotropic turbulence in
\cite{Ilya_Semion} and \cite{galperin}, respectively.
In the present work, we compare $\nu (k \vert k_c)$ for the inverse
energy cascade regime calculated from DNS data with those predicted by
TFM and the RG theory.

We solve Eq. (\ref{one}) numerically in a periodic box of the
size $2 \pi \times 2 \pi$ using $512^2$ Fourier modes.
The numerical scheme involves a Fourier-Galerkin pseudo-spectral spatial
approximation with implicit Adams-type second order stiffly stable
time-stepping scheme \cite{Stiffly_stable}. In order to increase the effective
inertial range, mode-selective hyperviscosity \cite{Benzi} of the
form $\nu(k)=\nu_L(k)+\nu_S(k)=A_L k^{-10}+A_S k^{14}$ has been introduced
in the vorticity equation (\ref{one}) instead of the molecular viscosity.
The constant coefficients $A_L$ and $A_S$ have been selected empirically
so as to minimize distortion of the energy inertial subrange.

To simulate the inverse energy cascade, high wave number forcing
$$
f(k,t)=A_f \left ( \delta_{k,k_f-1} + \delta_{k,k_f} + \delta_{k,k_f+1}
\right ) e^{i\phi(t)}
$$
is introduced in the vorticity equation; here, $A_f$ is the forcing
amplitude, and random variable $\phi(t)$ is a white-noise function
of time $t$ uniformly distributed in the interval $[0, 2 \pi]$.
The results are not sensitive to initial vorticity field
distributions, including the extreme case of the zero field.
A series of numerical experiments with various $k_f$ and
other flow parameters have been performed.  Here we shall report only the
results with $k_f = 98$ since they gave the broadest
inertial range.  Other parameters used in these simulations were
$A_f = 0.03244, ~A_L = 0.5$, and $A_S = 0.95 \times 10^{-34}$.  The value of
$A_S$ chosen is somewhat high, to ensure efficient enstrophy dissipation
and to eliminate the need for the dealiasing.

In Figs 1a,b we plot the total energy
$E_{{\rm tot}}(t)=\int_0^{+\infty}k^{-2}\Omega(k,t)dk$ and enstrophy
$\Omega_{{\rm tot}}(t)=\int_0^{+\infty}\Omega(k,t)dk$ as functions of time.
In Fig. 1a, one can see that the energy grows with time and eventually
tends to reach a steady state. However, the drift towards the energy steady
state is significantly slower than towards that of the enstrophy.
Defining the rms velocity as $V_{\rm{rms}}^2=\sum_{\bf k}\vert
{\bf u(k)}\vert^2$ and the characteristic turnover time of the largest eddies
as $\tau_{tu}=2 \pi / V_{\rm {rms}}$, we infer from Fig. 1b that
a steady state for the total enstrophy was achieved after about
$1.2$ $\tau_{tu}$, while about $5 \tau_{tu}$ were required to attain a
steady state for the total energy.  Note however that all the modes with
$k > 5$ have reached the steady state after $t \approx 2 \tau_{tu}$, and
only the largest modes were still developing.  The results presented below
pertain to the integration time $t \le 10^4$, before the energy saturates
at low wave numbers.

In Fig. 2 we plot the time-averaged energy spectrum obtained after about
$5$ $\tau_{tu}$. The inertial range $E\propto k^{-x}$ extends over more
than a decade in wave number space. Mean square line-fitting over the
interval $k \in (12, 50)$ gives the scaling exponent close to the Kolmogorov
value of ${5 \over 3}$.  Note that good agreement with the Kolmogorov scaling
in the energy sub-range has been reported recently in \cite{maltrud93} for
$256^2$ simulations and in \cite{leslie} for very high resolution simulations
with $2048^2$ Fourier modes. In Fig. 2 we also plot a compensated energy
spectrum,
$k^{5/3} \epsilon^{-2/3} E(k)$, where $\epsilon$
is the energy transfer rate.  The value of the Kolmogorov constant
calculated from this data is about $C_k=6.2$, in reasonable agreement
with $5.8$, calculated from DNS in \cite{Maltrud_Vallis} using the
$256^2$ resolution and $6.69$, obtained analytically in
\cite{Kraichnan_71} on the basis of TFM.

In Figs 3a,b we plot the $k$-dependent energy and enstrophy flux functions,
defined as $\Pi_E(k)=\int_0^kT_{\Omega}(n)n^{-2}dn$ and
$\Pi_{\Omega}(k)=\int_0^kT_{\Omega}(n)dn$, respectively. As expected, an
inverse energy cascade with constant energy transfer rate
$\epsilon$ develops for $k < k_f=98$, see Fig. 3a.
For $k > k_f$, $\Pi_E(k)$ and $\Pi_{\Omega}(k)$ both quickly fall to zero,
due to the strong dissipation at wave numbers adjacent to $k_f$.  In
Fig. 3b, we see that the flux of enstrophy in the energy sub-range,
$k < k_f$, is zero.  Strong enstrophy flux is observed for
$k > k_f$, until the enstrophy dissipation takes over and suppresses
the flux of enstrophy into even smaller scales. The resolution employed
in this study was insufficient to detect a well defined enstrophy
transfer range. The results plotted in Figs 3a,b also indicate that the
numerical scheme used conserves both total energy and enstrophy, since
$\Pi_{\Omega}(0)=\Pi_{\Omega}(\infty)=0$ and $\Pi_E(0)=\Pi_E(\infty)=0$.

We have calculated $k$ and $k_c$-dependent enstrophy transfer function
$T_{\Omega}({ k} \vert k_c)$ employed in (\ref{four})
by computing the third-order vorticity cumulants in (\ref{three})
extending the integration only over those ${\bf p}$ and ${\bf q}$ that
either $p \ge k_c$ or $q \ge k_c$. We have set $k_c=50$, well inside the
energy inertial subrange.

The DNS-inferred normalized TPEV [ viz., the function
$N(k / k_c)= \nu(k \vert k_c) / \vert \nu(0 \vert k_c) \vert$] is plotted in
Fig. 4, along with the TFM-based \cite{Kraichnan_76} and RG-based
\cite{Ilya_Semion} analytical predictions. The agreement between the
DNS-based results and the TFM and RG theories is very good over the entire
energy transfer range, up to the wave numbers close to $k_c$, where the
DNS data saturates, while TFM and RG curves exhibit sharp cusp.  The physics
leading to this cusp is as follows.  As closer $k$ approaches $k_c$, as more
elongated triads with either $p$ or $q \ll k_c$ become involved in the
energy exchange between the mode $k$ and the subgrid scale modes.
The contribution from these triads results in
the cusp behavior of the theoretical TPEV. However, in finite box DNS
with large-scale energy removal, the energy of small wave number modes
is reduced (see Fig. 2), which implies that instead of the sharp growth,
the TPEV should saturate at $k \to k_c$.  To illustrate
and quantify this explanation, we recalculated the RG-based TPEV
with the enstrophy spectrum in (\ref{five}) corrected at $k \le 5$
according to Fig. 2.  In Fig. 5, we compare the DNS- and RG-based
TPEV in their actual values, whereas the RG calculations were based upon
the value of $\epsilon$ found from DNS.  The agreement
between the two is very good. We have also calculated TPEV for $k_c = 35, 45$
and 55 and found that the DNS-inferred TPEV scales with $k_c^{-4/3}$, in
full agreement with the Kolmogorov and Richardson laws.  At all values of
$k_c$ tested an equally good agreement between DNS data and RG
predictions was observed.

The good agreement demonstrated in Figs. 4 and 5 provides an indirect
validation of TFM and RG results for isotropic 2D turbulence.

\acknowledgments

Authors would like to thank Eric Jackson for valuable help with some
programming issues, and Robert Kraichnan who kindly provided his data for
TFM-based eddy viscosity. This research has been partially supported by
ONR Grants N00014-92-J-1363, N00014-92-C-0089 and N00014-92-C-0118,
NSF Grant OCE 9010851, and the Perlstone Center for the Aeronautical
Engineering Studies.  The computations were performed on Cray Y-MP of
NAVOCEANO Supercomputer Center, Stennis Space Center, Mississippi.

\begin{figure}
\caption{Evolution of the total energy $E(k)$ (dotted line) and enstrophy
$\Omega(k)$ (solid line) towards the steady state.  Dashed line denotes the
total energy with the first 6 modes excluded. }
\label{fig1}
\end{figure}

\begin{figure}
\caption{Energy spectrum $E(k)$ (solid line) and compensated
energy spectrum $E(k) k^{5/3} \epsilon^{-2/3}$ (dotted line).}
\label{fig2}
\end{figure}

\begin{figure}
\caption{The energy flux $\Pi_E (k)$ (solid line) and the enstrophy flux
$\Pi_{\Omega} (k)$ (dotted line).}
\label{fig3}
\end{figure}

\begin{figure}
\caption{Normalized two-parametric eddy viscosity from DNS (dots),
from TFM (dashed line), and from RG (solid line).}
\label{fig4}
\end{figure}

\begin{figure}
\caption{Actual two-parametric eddy viscosity from DNS (dots) and from RG
(solid line). In RG calculations, the energy spectrum for $k < 5$ was corrected
in accordance with the DNS results, Fig. 2.}
\label{fig5}
\end{figure}

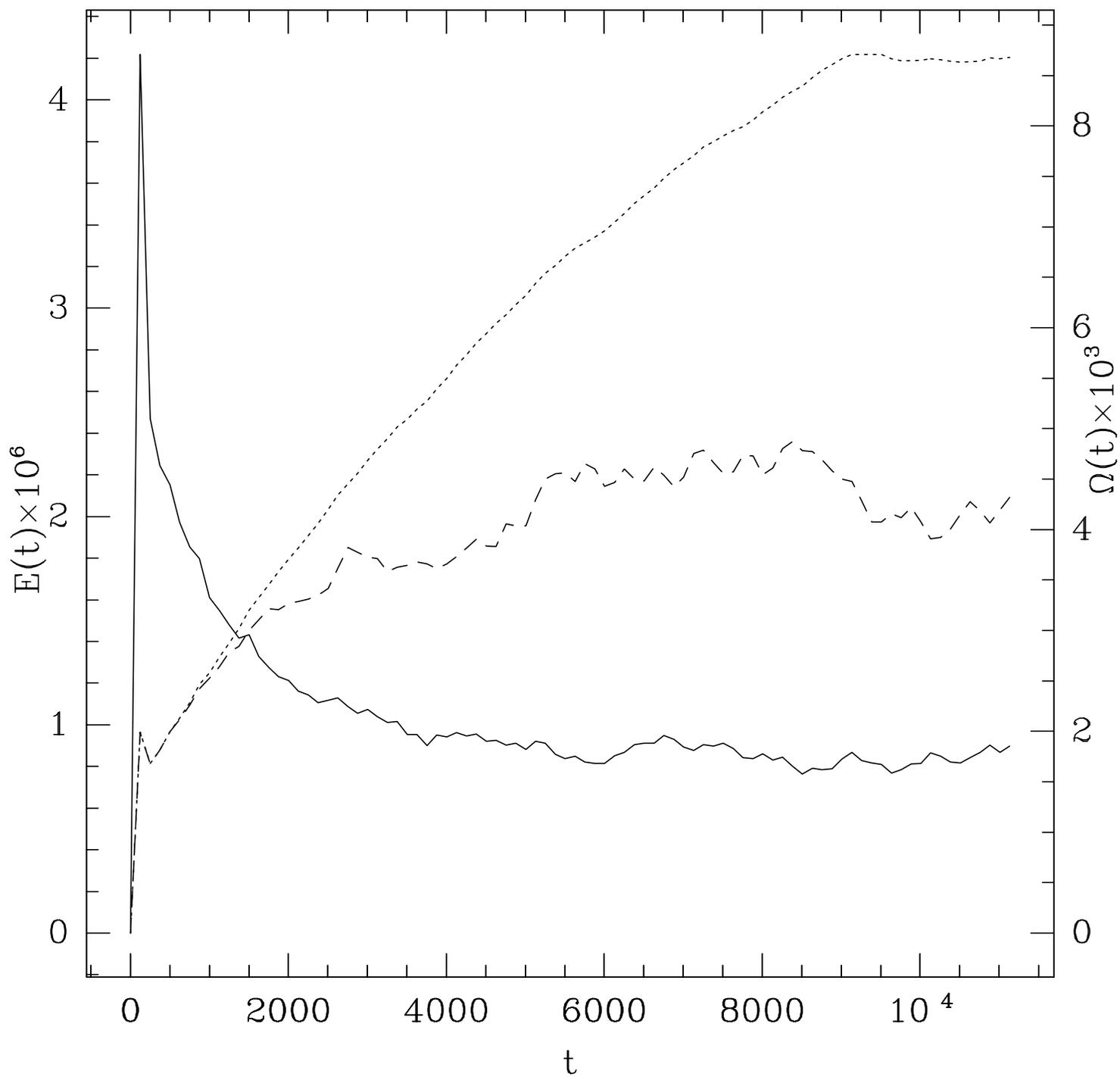
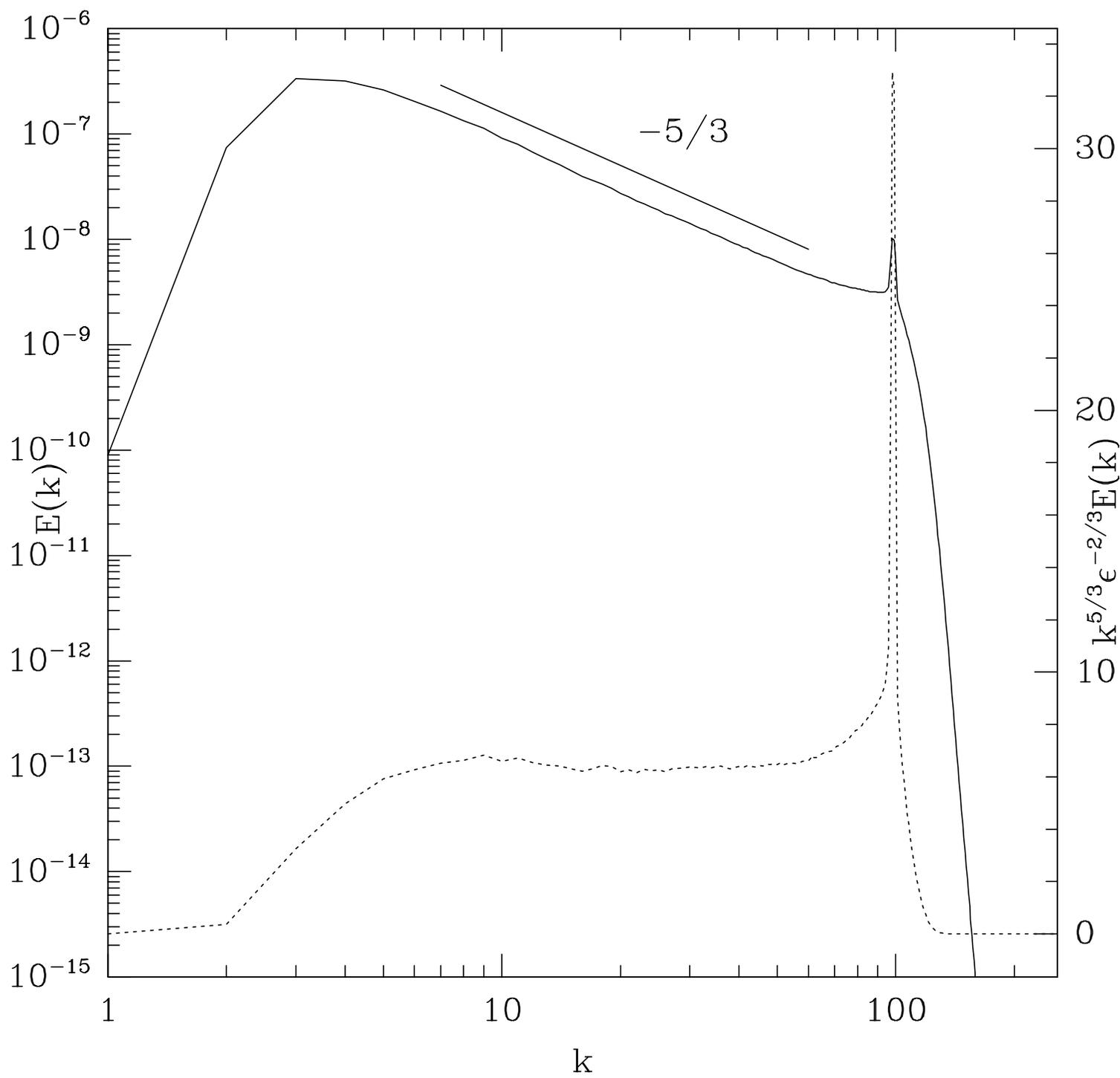
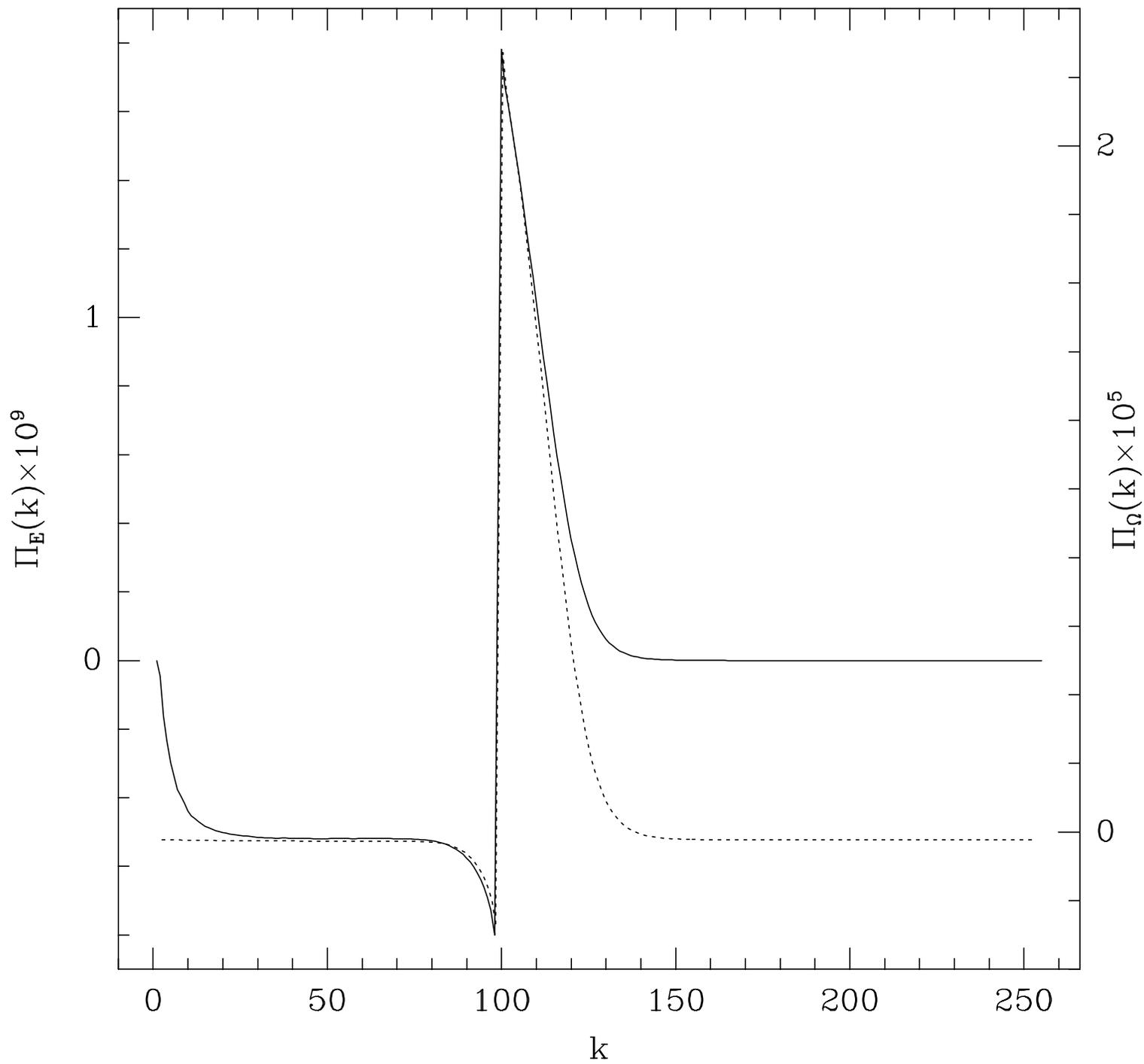
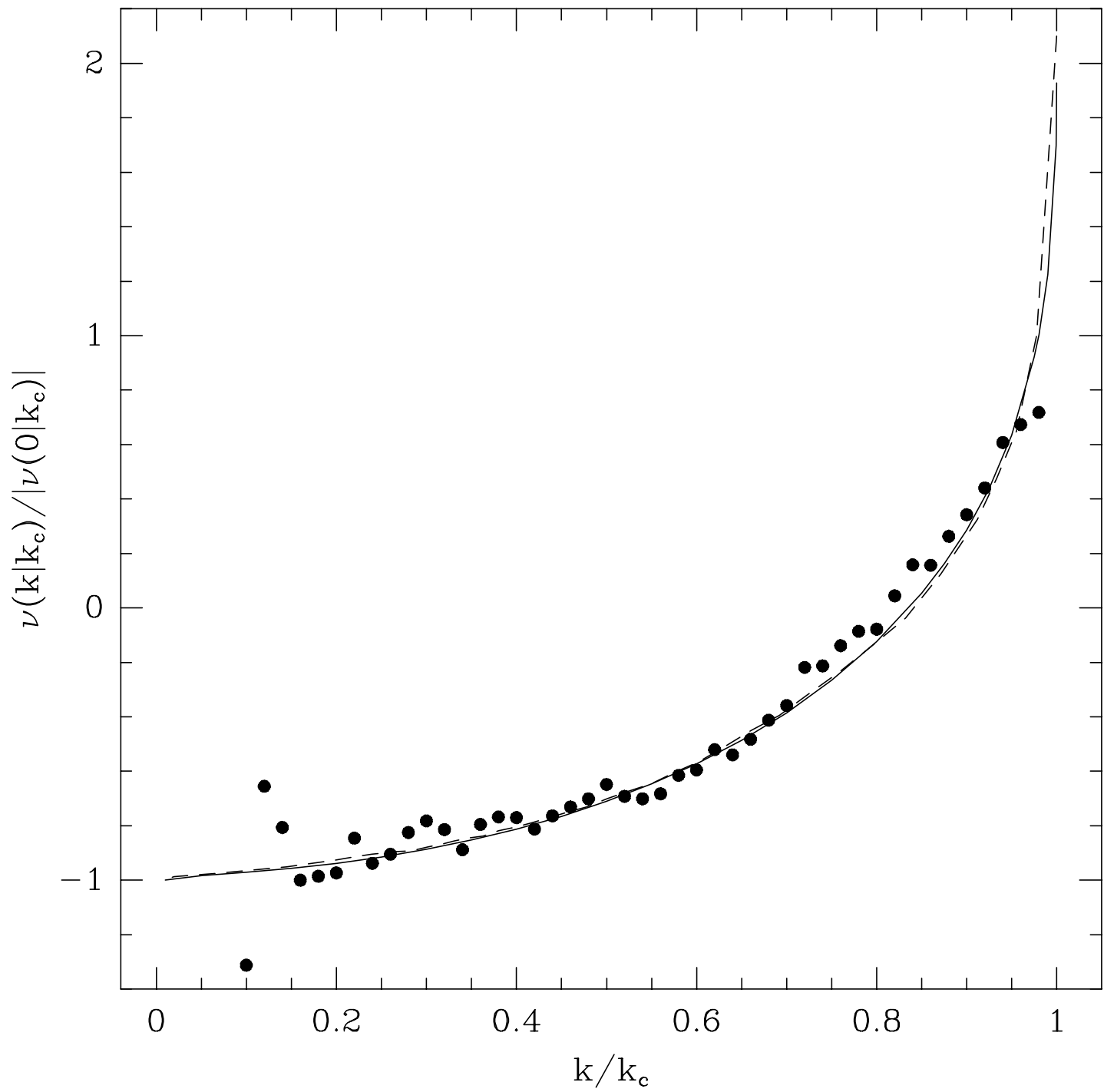
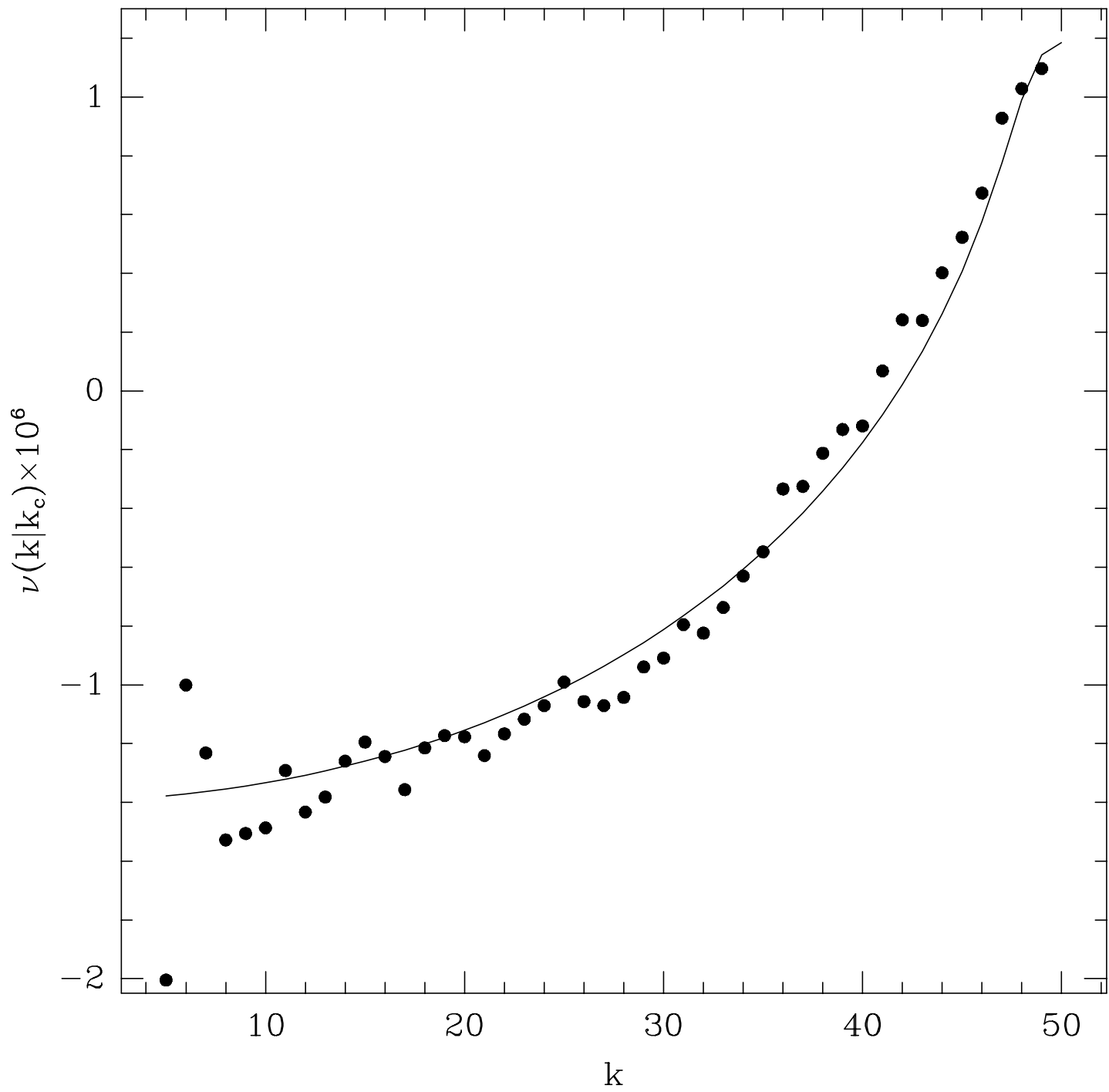
\end{document}